\documentclass[aps,prl, twocolumn,superscriptaddress,nofootinbib]{revtex4}
\usepackage{amsmath}
\usepackage{graphicx}
\usepackage{amssymb}
\usepackage[colorlinks,citecolor=blue]{hyperref}

\newcommand{\beq}{\begin{equation}}
\newcommand{\eeq}{\end{equation}}
\newcommand{\bea}{\begin{eqnarray}}
\newcommand{\eea}{\end{eqnarray}}
\newcommand{\met}{\not{\!\!{\rm E}}_{T}}
\newcommand{\nn}{\nonumber}

\begin{document}

\title{Measuring Trilinear Higgs Coupling in $WHH$ and $ZHH$ Productions at the HL-LHC}

\author{Qing-Hong Cao}
\email{qinghongcao@pku.edu.cn}
\affiliation{Department of Physics and State Key Laboratory of Nuclear Physics and Technology, Peking University, Beijing 100871, China}
\affiliation{Collaborative Innovation Center of Quantum Matter, Beijing, 100871, China}
\affiliation{Center for High Energy Physics, Peking University, Beijing 100871, China}

\author{Yandong Liu}
\email{ydliu@pku.edu.cn}
\affiliation{Department of Physics and State Key Laboratory of Nuclear Physics and Technology, Peking University, Beijing 100871, China}

\author{Bin Yan}
\email{binyan@pku.edu.cn}
\affiliation{Department of Physics and State Key Laboratory of Nuclear Physics and Technology, Peking University, Beijing 100871, China}

\begin{abstract}
Determination of trilinear Higgs coupling ($\lambda_{HHH}=\kappa\lambda_{HHH}^{\rm SM}$) through Higgs pair productions is a major motivation for the LHC high luminosity phase. We perform a detailed collider simulation to explore the potential of measuring $\lambda_{HHH}$ in the $VHH$ ($V=W,Z$) production at the HL-LHC. We find that the trilinear Higgs coupling in the SM ($\lambda_{HHH}^{\rm SM}$) could be measured at the level of $1.3\sigma$. Combining with the gluon fusion, vector boson fusion and $t\bar{t}HH$ channels, $\lambda_{HHH}^{\rm SM}$ is expected to be measured at the level of $3.13\sigma$. If no evidence of Higgs pair productions were observed, the $VHH$ production, together with the gluon fusion channel, would impose a bound of $0.5\leq \kappa \leq 2.2$ at the 95\% confidence level. 
\end{abstract}

\maketitle

\section{Introduction}

\begin{figure}
\includegraphics[scale=0.32]{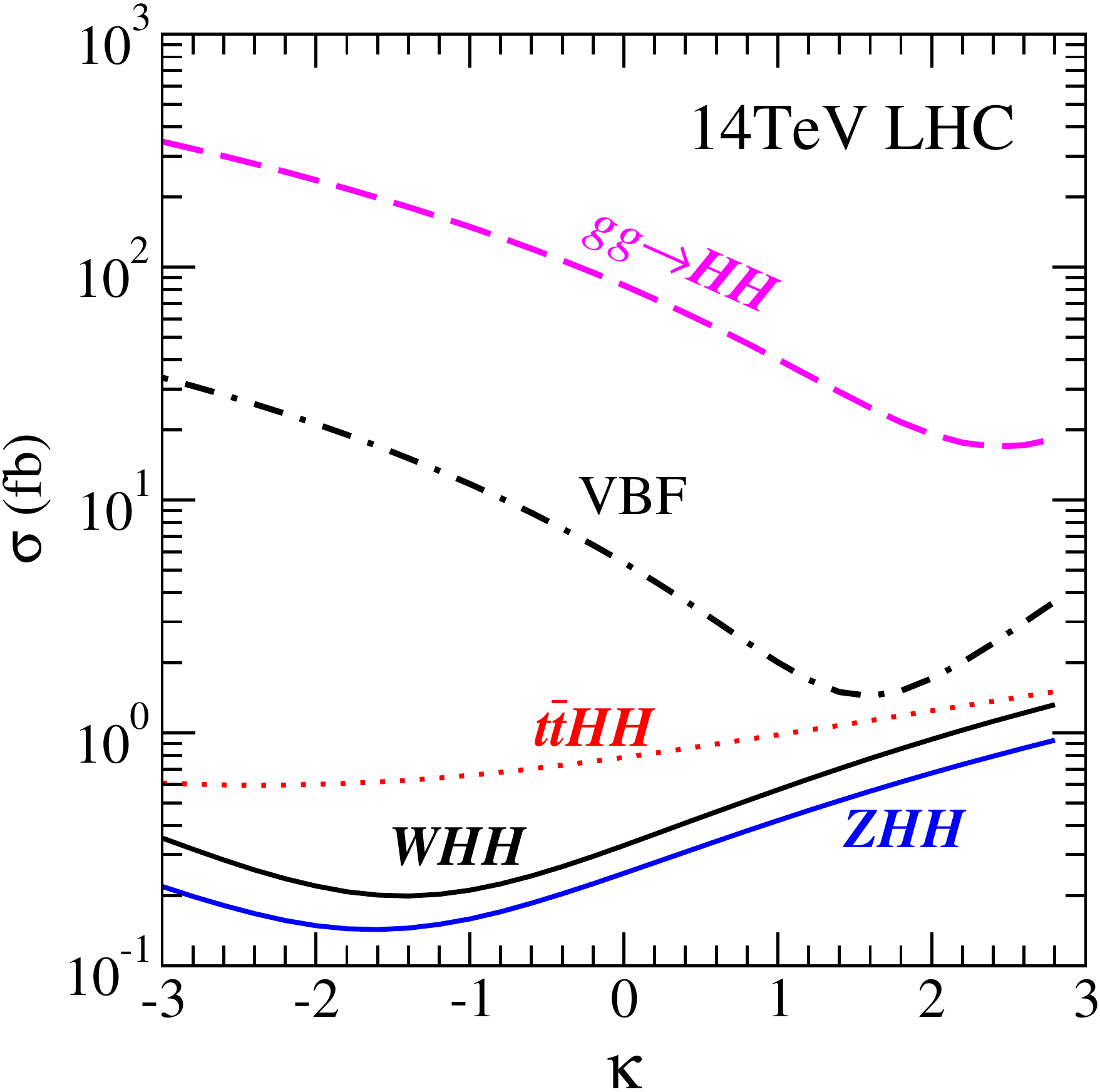}
\caption{The production cross section (in unit of fb) of Higgs boson pair production as a function of $\kappa$ at the 14 TeV LHC.}
\label{fig:xsec}
\end{figure}

The trilinear Higgs  coupling ($\lambda_{HHH}$) is an important parameter of the Higgs boson potential in the Standard Model (SM).  Even though $\lambda_{HHH}$ is related to the Higgs boson mass ($m_H$) in the SM, it might deviate from the SM value $\lambda_{HHH}^{\rm SM}$ in new physics (NP) models~\cite{Kanemura:2002vm,Han:2003wu,Kanemura:2004mg,Bhattacherjee:2014bca,Barger:2014qva,Hespel:2014sla,Wu:2015nba}. For example, the deviation of the trilinear Higgs coupling might emerge from non-vanishing higher-dimension operators starting with dimension 6.  It is then of critical importance to measure $\lambda_{HHH}$ to test the SM. Such a goal will only be successful if information from a range of production channels of Higgs boson pairs is included. There are five major channels of Higgs pair productions. Figure~\ref{fig:xsec} plots the inclusive cross sections of all those five channels as a function of $\kappa$ at the 14~TeV Large Hadron Collider (LHC), where the factor $\kappa$ is introduced to describe possible NP effects in the trilinear Higgs coupling as $\lambda_{HHH}\equiv \kappa \lambda_{HHH}^{\rm SM}$. The leading production channel is the so-called gluon fusion (GF) channel, $gg\to HH$~\cite{Glover:1987nx,Moretti:2004wa,Dawson:2012mk,Baglio:2012np,Papaefstathiou:2012qe,Dolan:2012rv,Shao:2013bz,Goertz:2013kp,Li:2013flc,
Chen:2014xwa,Chen:2014xra,Frederix:2014hta,Maltoni:2014eza,Li:2015yia,Dawson:2015oha,He:2015spf,Cao:2015oaa,deFlorian:2013jea,
deFlorian:2015moa,Dawson:1998py,deFlorian:2013uza,Grigo:2013rya,Grober:2015cwa,Grigo:2014jma,Azatov:2015oxa,
Contino:2012xk,Plehn:1996wb,Nishiwaki:2013cma,Slawinska:2014vpa,Barr:2013tda,Barger:2013jfa,deLima:2014dta,Wardrope:2014kya}, the subleading channel is the vector boson fusion (VBF) process, $qq \to qqHH$~\cite{Moretti:2004wa,Baglio:2012np,Dolan:2013rja,Liu-Sheng:2014gxa,Frederix:2014hta,Dolan:2015zja}, the third channel is $t\bar{t}HH$ production~\cite{Moretti:2004wa,Baglio:2012np,Englert:2014uqa,Liu:2014rva,Frederix:2014hta}, while the last two channels are $WHH$ and $ZHH$ productions~\cite{Barger:1988jk,Moretti:2004wa,Baglio:2012np,Frederix:2014hta}. The GF channel has drawn a lot of attentions owing to its large cross section. Searches for Higgs pairs in the decay modes of $bb\bar{b}\bar{b}$, $b\bar{b}\tau\tau$, $b\bar{b}WW$ and $\gamma\gamma b\bar{b}$ have been carried out by the ATLAS collaboration recently~\cite{Aad:2014yja, Aad:2015uka, Aad:2015xja} and a combined upper limit of $\sigma(gg\to HH)\leq 0.69~{\rm pb}$ is observed. On the other hand, the VBF channel is shown to be less promising to measure the trilinear Higgs coupling~\cite{Dolan:2015zja} as it cannot compete with the process of $gg\to HHqq$, i.e. high order QCD  corrections to the GF channel. The potential of probing $\lambda_{HHH}$ in the $t\bar{t}HH$ channel at the high-luminosity (HL) LHC (a 14~TeV $pp$ collider with an integrated luminosity of $3000~{\rm fb}^{-1}$) is examined in Refs.~\cite{Englert:2014uqa,Liu:2014rva} which shows that a bound of $\kappa\leq 2.51$ can be reached at 95\% CL. Unfortunately, the $WHH$ and $ZHH$ productions are not yet studied carefully in the literature, which is not surprising because the $VHH$ production rate are quite small. In this work we focus on the $VHH$ ($V=W^\pm,Z$) production channel and demonstrate that, the $VHH$ production could probe the trilinear Higgs coupling at a comparable level as the GF and $t\bar{t}HH$ channels at the HL-LHC.

The $VHH$ production has many advantages over others production channels. First, the charged lepton and invisible neutrino from $W$- or $Z$-boson decays in the $VHH$ production provide a good trigger of signal events. 
Note that $\sigma(VHH)$ is about one tenth of $\sigma(gg\to HH)$. However, after including the branching ratio (BR) of $V$-boson and Higgs boson decays, the cross section of $VHH$ channel is comparable to the GF channel with subsequent decays $HH\to \gamma\gamma b\bar{b}$.  For example, 
\begin{align}
& \sigma(W^\pm HH)\times {\rm BR}(W^\pm \rightarrow \ell^\pm \nu_\ell,HH\rightarrow b\bar{b}b\bar{b})=0.088\rm{fb},\nn\\
& \sigma(ZHH)\times {\rm BR}(Z\rightarrow \nu\bar{\nu},HH\rightarrow b\bar{b}b\bar{b})=0.059\rm{fb}, \nn\\
& \sigma(gg\rightarrow HH)\times {\rm BR}(HH\rightarrow\gamma\gamma b\bar{b})=0.15\rm{fb},\nn
\end{align}
where $\ell=e,\mu$. Here, we use the branching ratios at the tree level: ${\rm Br}(H\to b\bar{b})=0.83$, ${\rm Br}(Z\to\nu\bar{\nu})=0.20$, and ${\rm Br}(W^+\to \ell^+\nu_\ell)=0.11$. Also, the SM backgrounds can be dramatically reduced by tagging the charged lepton and missing neutrino from the $V$-boson decay.

Second, all the five channels of Higgs pair productions do not only involve a diagram with $\lambda_{HHH}$ but also additional contributions which then dilute the sensitivity to $\lambda_{HHH}$. The GF channel can be modified by several NP effective operators~\cite{Cao:2015oaa} which might arise from colored NP particles inside the loop diagrams or  from the modified top-quark Yukawa coupling, etc. The $VHH$ production could be modified by the $HVV$ anomalous couplings, which are tightly bounded~\cite{CMS:2015kwa,ATLAS-CONF-2015-044}.

Last but not least, as depicted in Fig.~\ref{fig:xsec}, the dependence of $\sigma(VHH)$ on $\kappa$ is different from those of the GF,  VBF and $t\bar{t}HH$ channels. The $t\bar{t}HH$ production is dominated by the large top Yukawa coupling such that it has a mild dependence on $\kappa$. On the other hand, both the GF and the VBF channels are sensitive to a negative $\kappa$ while the $VHH$ channel is sensitive to a positive $\kappa$. One can use the $VHH$ production to probe positive $\kappa$ and the GF channel to limit negative $\kappa$. 

The $\kappa$ dependence can be understood as follows. 
The GF channel exhibits a large cancellation between the triangle diagram and the box diagram around the threshold of Higgs boson pairs~\cite{Glover:1987nx}. When $\kappa<0$ the two diagrams interfere constructively so as to enhance the cross section.

\begin{figure}
\includegraphics[scale=0.3]{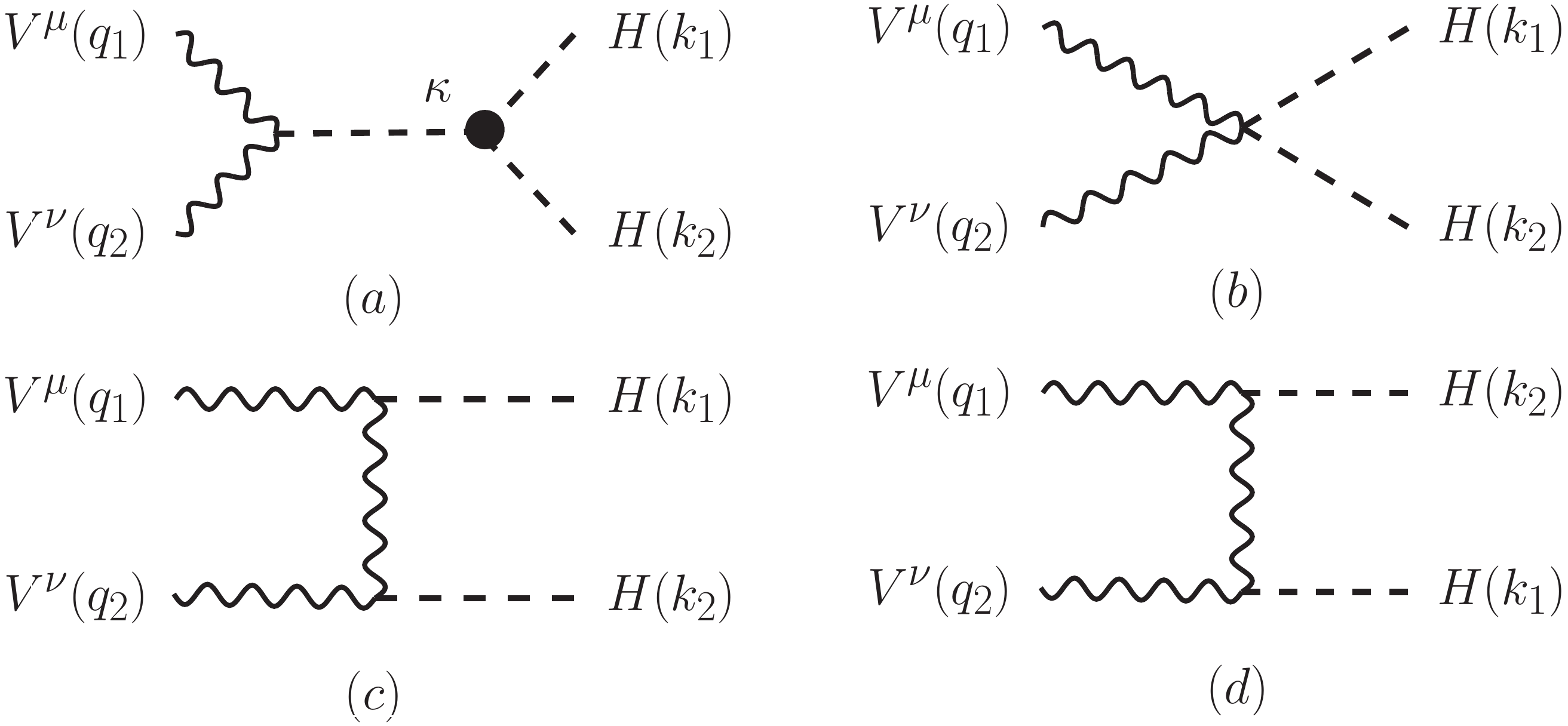}
\caption{Feynman diagrams of $V^\mu V^\nu \to HH$.}
\label{fig:VVHH}
\end{figure}

The VBF and $VHH$ channels share the same subprocess of $V^\mu V^\nu \to HH$ and are related to each other by crossing symmetry. Consider the VBF channel first. The matrix element of $V^\mu(q_1) V^\nu(q_2) \to H(k_1) H(k_2)$ is 
\bea
M^{\mu\nu}&=& g^{\mu\nu}\left[ \kappa \frac{m_V^2}{v^2}\frac{6m_H^2}{\hat{s}-m_H^2}+\frac{2m_V^2}{v^2} \right. \nn\\
& +& \left. \frac{4m_V^4}{v^2}\left(\frac{1}{\hat{t}-m_V^2} +\frac{1}{\hat{u}-m_V^2}\right)\right] +{\rm others},
\label{eq:VVHH}
\eea
where $m_V$ denotes the mass of $V$-boson and $q_{1,2}$ ($k_{1,2}$) denotes the momentum of the $V$ (Higgs) boson, respectively. Figure~\ref{fig:VVHH} shows Feynman diagrams of $V^\mu V^\nu \to HH$.
For the VBF channel, $\hat{s}=(q_1+q_2)^2$, $\hat{t}=(q_1-k_1)^2<0$ and $\hat{u}=(q_1 - k_2)^2<0$. Near the threshold of Higgs boson pairs, $\hat{s}\sim 4m_H^2$ and $\hat{t}\simeq \hat{u}\sim 0$. It gives rise to 
\beq
M^{\mu\nu} \sim \frac{2m_V^2}{v^2}(\kappa - 3) g^{\mu\nu} + \cdots, \nn
\eeq
yielding a small cross section around $\kappa \sim + 3$ and a large cross section for $\kappa<0$. The sub-amplitude of $VHH$ production can be obtained from $VV\to HH$ by crossing one gauge boson from initial state to final state.  In the vicinity of the thresholds of $HH$ and $VH$ pairs, 
$\hat{s}\sim 4m_H^2$ and $\hat{t}=\hat{u}\sim (m_H+m_V)^2$.
That yields
\beq
M^{\mu\nu}\sim \frac{2m_V^2}{v^2}\left(\kappa + 1 +\frac{4m_V^2}{m_H(m_H+2m_V)}\right) g^{\mu\nu} + \cdots,\nn
\eeq
which leads to a small cross section around $\kappa \sim -2$ and a large cross section for $\kappa>0$. 

Next we perform a collider simulation at the parton level to investigate the sensitivity of the HL-LHC on the $\lambda_{HHH}$ measurement through the $VHH$ production. 

\section{The $WHH$ production}

Our signal consists of both $W^+HH$ and $W^-HH$ productions. The signal and background events are generated with MadEvent~\cite{Alwall:2011uj}. To include higher order QCD corrections, we multiply the cross section of the signal at the tree-level with a factor of $K_{WHH}=1.39$~\cite{Baglio:2012np}. As the production rate is quite small, we demand both the Higgs bosons decay into the $b\bar{b}$ pair which has the largest decay branching ratio among all the decay modes of the Higgs boson. In order to trigger the signal events, we demand a leptonic decay  of the $W$-boson which gives rise to a charged lepton and an invisible neutrino in the final state. The event topology of our signal is characterized by one isolated charged lepton ($\ell^\pm$), four $b$-jets and a large missing transverse momentum ($\met$)  from the missing neutrino. Both electrons and muons are used in our analysis. 

In the article the QCD corrections to the signal processes are taken into account by introducing a constant $K$ factor. It is worthwhile discussing how much our result will be influenced by the QCD corrections. The NNLO QCD corrections to the $VHH$ productions is calculated in Refs.~\cite{Baglio:2012np}, which shows the errors of cross sections are dominated by PDF uncertainty. For example, it shows that the total uncertainty is $+3.7\%$ and $-3.1\%$ for $WHH$ production, $7.0\%$ and $-5.5\%$ for $ZHH$ production at the 14 TeV LHC. The fully differential cross section of $WHH$ production at the next-to-next-to-leading order is calculated in Ref.~\cite{Li:2016nrr}. It shows the QCD effects mildly modify the transverse momentum distributions of $W$-boson and the Higgs bosons, which do not alter the acceptance of kinematic cuts used in this study. 

\begin{table*}
\caption{Numbers of $WHH$ signal ($\kappa=1$) and background events after a series of cuts which are applied sequentially at the 14 TeV LHC with an integrated luminosity of $3000~{\rm fb}^{-1}$.
}
\label{tbl:whhcut}
\begin{tabular}{l|c|c|c|c|c|c|c|c}
\hline
&$WHH$&$Wbb\bar{b}\bar{b}$&$Zbb\bar{b}\bar{b}$&$t\bar{t}$&$t\bar{t}j$ &$t\bar{t}H(\rightarrow b\bar{b})$&$t\bar{t}Z(\rightarrow b\bar{b})$&$t\bar{t}b\bar{b}$ \\
\hline
Basic cuts     & 200.9   &157770   &266580&$4.26\times 10^8$&$1.0\times 10^9$&296716&97888&$7.0\times10^6$ \\
\hline
Selection cuts &34.6    &544.4   &44.9&$3.62\times 10^7$&$3.8\times 10^7$&1454.0&442.4&65590.7\\
\hline
$4b$ tagging   &7.2     &97.2    &9.9&767.6&1002&265.1&71.6&922.3\\
\hline
$\chi_{HH}<1.6$&7.2     & 2.3    &0&170.6&0&45.8&0.1&51.8  \\
\hline 
$\chi_{tt}>3.2$ &4.8    & 0.6    &0&0&0&20.5&0.05&47.9\\
\hline
$m_T~\&~H_T$ cuts&3.5 &0.2     &0&0&0&4.1&0&2.0\\
\hline
\end{tabular}
\end{table*}

The SM backgrounds are rather complicated. In order to reduce the huge SM backgrounds, we require all the four hard jets are tagged as $b$-jets. In the study we also take into account the possibility that a light quark jet fakes a $b$-jet, with mistag efficiency $\epsilon_{j\to b}=0.1\%$ with $j=u,d,s$ and $\epsilon_{c\to b}=10\%$. 
The major SM backgrounds include $W^\pm(\to \ell^\pm\nu_\ell)bb\bar{b}\bar{b}$, $Z(\to \ell^+ \ell^-)bb\bar{b}\bar{b}$,
$t\bar{t},~t\bar{t}j$, $t\bar{t}H$, $t\bar{t}Z$ and $t\bar{t}b\bar{b}$. 
To mimic the signal events, top-quark pairs in the $t\bar{t}$ and $t\bar{t}j$ backgrounds are required to decay into semi-leptonic final states while top-quark pairs in the $t\bar{t}H$, $t\bar{t}Z$ and $t\bar{t}b\bar{b}$ backgrounds decay into dilepton final states. 
To take into account higher order QCD corrections, we multiply the $t\bar{t}H$ and $t\bar{t}Z$ backgrounds with a factor of $K_{t\bar{t}H}=1.22$~\cite{Beenakker:2001rj,Beenakker:2002nc,Dawson:2002tg,Dawson:2003zu,Yu:2014cka,Frixione:2014qaa,Maltoni:2015ena,Frixione:2015zaa} and $K_{t\bar{t}Z}=1.49$~\cite{Lazopoulos:2008de,Garzelli:2011is,Kardos:2011na,Garzelli:2012bn,Maltoni:2015ena,Frixione:2015zaa}, respectively. 

When generating both the signal and background events, we impose {\it basic} cuts as follows: $p_T^{\ell^\pm,b,j} > 5~{\rm GeV}$ with $|\eta^{\ell^\pm,b,j}| <5$, where $p_T$ and $\eta$ denotes the transverse momentum and rapidity, respectively. To get the isolated objects, we require the cone distance $\Delta R_{mn}\equiv \sqrt{(\eta^m-\eta^n)^2 + (\phi^m-\phi^n)^2}$ between the object $m$ and $n$ is at least 0.4. The numbers of signal and background events are shown in the second row of Table~\ref{tbl:whhcut}.

There are also many other SM backgrounds involving light non-$b$ jets, e.g. $Wb\bar{b}c\bar{c}$ (5.6~fb), $Wcc\bar{c}\bar{c}$ (3.8~fb), $Wb\bar{b}jj$ ($\sim 10^3~\rm{fb}$), $Wc\bar{c}jj$ ($\sim 10^3~\rm{fb}$), $Wjjjj$ ($\sim 10^5~\rm{fb}$), $WWZ$ (95.9~fb), $ZZZ$ (10.4~fb), $WZZ$ (30.2~fb), $WWW$ (12.6~fb) and $t\bar{t}ZZ$ ($1.8~\rm{fb}$). The numbers shown inside the bracket denote the production cross section after imposing $p_T^{b,j}\geq40$ GeV with $|\eta^{b,j}|\leq2.5$. All the above light-jet backgrounds can be safely ignored after tagging four $b$-jets. For example, $\sigma(WWZ\to \ell^\pm bbbb+\met) \sim 10^{-5}~\rm{fb}$, $\sigma(ZZZ\to \ell^\pm bbbb+\met) \sim 10^{-3}~\rm{fb}$. From now on we ignore those light-jet backgrounds in our collider simulations.

At the analysis level, all the signal and background events are required to pass a set of {\it selection} cuts~\cite{Aad:2015uka}: 
\begin{align}
& p_T^e \geq 15~{\rm GeV}, && p_T^\mu \geq 10~{\rm GeV}, && p_T^b \geq 40~{\rm GeV},\nn\\
& \left| \eta^{e,\mu,b} \right|\leq 2.5~,&& \Delta R_{bb,b\ell}>0.4, && \met\geq 40~{\rm GeV}.
\label{eq:cut}
\end{align}
We model detector resolution effects by smearing the final state energy according to $\delta E/E= \mathcal{A}/\sqrt{E/{\rm GeV}}\oplus \mathcal{B}$, where we take $\mathcal{A}=10(85)\%$ and $\mathcal{B}=1(5)\%$ for leptons(jets). We demand only one charged lepton and four hard jets in the central region of the detector. Those reducible backgrounds with more jets or charged leptons could mimic the experimental signature of the signal events if the $p_T$ of additional jets or leptons is less than 10~GeV or its  rapidity (in magnitude) is larger than 3.5~. 
As shown in the third row of Table~\ref{tbl:whhcut}, roughly $1/6$ of the signal events pass the selection cuts.

Once the four jets are trigged, one can require them to be $b$-jets. That significantly reduces the SM backgrounds consisting of light jets; see the fourth row of Table~\ref{tbl:whhcut}. The $b$-tagging efficiency depends on both $p_T^b$ and $\eta^b$. We adapt the $b$-tagging efficiency given in Ref.~\cite{Cao:2015oaa} which yields on average a $b$-tagging efficiency of 70\% in our analysis.

\begin{table*}
\caption{The numbers of $ZHH$ ($\kappa=1$) signal and background events after a series of cuts which are applied sequentially at the 14 TeV LHC with an integrated luminosity of $3000~{\rm fb}^{-1}$.
}
\begin{tabular}{l|c|c|c|c|c|c|c|c|c|c|c}
\hline
&$ZHH$   &$Zbb\bar{b}\bar{b}$  &$Zb\bar{b}c\bar{c}$ &$Zcc\bar{c}\bar{c}$  &$ZZ(\rightarrow b\bar{b})b\bar{b}$  &$t\bar{t}c\bar{c}$ &      $t\bar{t}b\bar{b}$& $t\bar{t}Z(\to b\bar{b})$ &$t\bar{t}H(\rightarrow b\bar{b})$&$t\bar{t}$&$t\bar{t}j$ \\
\hline
Basic cuts         &155.4   &$1.2\times10^6$    &$3.2\times10^6$ &$2.0\times 10^6$   & 24627   &$4.6\times10^6$   & $7.0\times10^6$ &97889 &296716&$4.3\times10^8$  &$1.0\times10^9$\\
\hline
Selection cuts	   &23.2   &2589.6    &4504.8   &2000.4     &308.0    &2131.3   &2435.0 &11.4 &29.4& $1.4\times10^6$  & $1.4\times10^6$ \\
\hline
$4b$ tagging       &4.8    &499.4     &19.2    &0           &56.6     &0      & 27.7  &2.0     &3.6& 28.4    &   66.8    \\
\hline
$\chi_{HH}<1.6$    &4.8    &10.8     &0       &0           &0.1      &0       & 0   &0.05    &0.7 &0&0  \\
\hline
\end{tabular}
\label{tbl:zhh}
\end{table*}

The four $b$-jets in the signal events originate from the Higgs boson decay. We first order the jets by their values of $p_T$ and then demand at least one combination of the four jets to be consistent with those expected for the $HH\to b\bar{b}b\bar{b}$ decay, i.e. 
\beq
\chi_{HH}\equiv \sqrt{ \left(\frac{m_{ij}-m_H}{\sigma_{m_H}}\right)^2+ \left(\frac{m_{i^\prime j^\prime}-m_H}{\sigma_{m_H}}\right)^2 } \le 1.6,
\label{eq:mh}
\eeq
where $m_{ij}$ denotes the invariant mass of the dijet $i$ and $j$, and $\sigma_{m_H}(= m_H/10)$ is the dijet mass resolution. All the signal events pass the cut while only 1\% of the background events remains. 

At this stage of analysis, the dominant background is from $t\bar{t}$  production. Following the ATLAS study~\cite{Aad:2015uka}, we check the compatibility with top-quark decay hypothesis with the following variable
\beq
\chi_{tt}=\sqrt{ \left(\frac{\tilde{m}_W- m_{W}}{\sigma_{m_W}}\right)^2+ \left(\frac{\tilde{m}_t - m_{t}}{\sigma_{m_t}}\right)^2 },
\eeq
with $m_W = 80.419~{\rm GeV}$ and $m_t = 173~{\rm GeV}$.
The $\sigma_{m_W}=0.1 m_W$ and $\sigma_{m_t}=0.1 m_t$ represent the dijet and three-jet system mass resolutions, $\tilde{m}_W$ and $\tilde{m}_t$ are the invariant mass of the $W$ and top candidates. If either dijet in an event has $\chi_{tt}\leq 3.2$ for any possible combination with an extra jet, the event is rejected. This requirement sufficiently reduces the $t\bar{t}$ background, whilst retaining $\sim 67\%$ of signal events; see the sixth row of Table~\ref{tbl:whhcut}. 

While the $\met$ in the signal events arises mainly from the missing neutrinos, that in the background events is contaminated by jets or leptons either falling into a large rapidity region or carrying a too small transverse momentum to be detected. To further suppress the SM backgrounds, we impose cuts on the transverse mass ($m_T$) of  $\met$ and charged lepton, $m_T(\ell^\pm, \met) = \sqrt{2 p_T^{\ell}\met\left(1-\cos\phi\right)}$ with $\phi$ being the azimuthal angle between $\ell^\pm$ and $\met$, and $H_T$ defined as the scalar sum of $p_T$'s of jets and charged lepton as follows:
\beq
m_T \leq m_W,~~H_T\geq 400~{\rm GeV}.
\eeq
The SM backgrounds are suppressed efficiently such that only 6.3 background events survive after all the cuts. We end up with 3.5 signal events.

Equipped with the optimal cuts shown above, we vary $\kappa$ to obtain a 5 standard deviations ($\sigma$) statistical significance using 
\beq
\sqrt{-2\left[(n_b + n_s) \log\frac{n_b}{n_s+n_b}+n_s\right]},
\label{eq:dis}
\eeq
from which we obtain a $5\sigma$ discovery significance requires $\kappa\geq 4.81$ or $\kappa\leq -7.68$. Here, $n_b$ and $n_s$ represents the numbers of the signal and background events, respectively. 
A discovery significance of the SM trilinear Higgs coupling ($\kappa=1$) is found to be around $1.29\sigma$, which is comparable to the projected significance derived from the GF channel by the ATLAS ($1.19\sigma$)~\cite{ATL-PHYS-PUB-2014-019} and CMS collaborations ($1.65\sigma$)~\cite{CMS:2015nat}.

In the case that no evidence of Higgs pair production is observed, one can set a $2\sigma$ exclusion limit on $\kappa$ from 
\beq
\sqrt{-2\left[n_b\log \frac{n_s + n_b}{n_b}-n_s\right]} = 2~,
\label{eq:ex}
\eeq
yielding $-5.11\leq\kappa\leq 2.24$~.

\section{The $ZHH$ Production}

Now consider the $ZHH$ production. We require that both the Higgs bosons decay into the $b\bar{b}$ pair and the $Z$ boson decays into neutrinos.
The topology of our signal events is characterized by four $b$-jets and a large $\met$ from the missing neutrinos. The major SM backgrounds are
\begin{align}
&Zbb\bar{b}\bar{b}, ~Zb\bar{b}c\bar{c}, ~Zcc\bar{c}\bar{c},~ZZ(\to b\bar{b})b\bar{b},~t\bar{t}c\bar{c}, \nn\\
&t\bar{t}b\bar{b},~t\bar{t}Z(\to b\bar{b}),~t\bar{t}H(\to b\bar{b}),~t\bar{t},~t\bar{t}j.\nn
\end{align}
Top quark pairs in the $t\bar{t}$ and $t\bar{t}j$ backgrounds are demanded to decay into semi-leptonic final states, while those pairs in the $t\bar{t}c\bar{c}$, $t\bar{t}b\bar{b}$, $t\bar{t}Z$ and  $t\bar{t}H$ backgrounds decay into both semi-leptonic and dilepton final states. The background $t\bar{t}Z$ with $Z\to \nu\bar{\nu}$ is negligible after requiring four $b$-tagged jets.  

The cross section of the signal is normalized to the NNLO precision~\cite{Baglio:2012np} while the $t\bar{t}H$ and $t\bar{t}Z$ backgrounds to the NLO accuracy~\cite{
Beenakker:2001rj,Beenakker:2002nc,Dawson:2002tg,Dawson:2003zu,Lazopoulos:2008de,
Garzelli:2011is,Kardos:2011na,Garzelli:2012bn,Yu:2014cka,Frixione:2014qaa,Maltoni:2015ena,Frixione:2015zaa}. We also consider a few SM backgrounds involving light jets, e.g. $ZZZ$ (10.4~fb), $WZZ$ (30.2~fb), $Zjjjj$ ($\sim 10^4~\rm{fb}$), $Zb\bar{b}jj$ ($\sim 10^3~\rm{fb}$), $Zc\bar{c}jj$ ($\sim 10^3~\rm{fb}$), $WZb\bar{b}$ (18~fb) and $t\bar{t}jj$ ($\sim 10^5~\rm{fb}$). The numbers shown inside the bracket denote the production cross section after imposing $p_T^{b,j}\geq40$ GeV with $|\eta^{b,j}|\leq2.5$ on the jets. Again we note that those light flavor jet backgrounds are negligible after requiring four $b$-tagged jets in the final state.

The {\it selection} cuts used in the $ZHH$ production are the same as those used in the $WHH$ channel (see Eq.~\ref{eq:cut}) except that now we demand $\met>100~{\rm GeV}$ to trigger the events. Table~\ref{tbl:zhh} displays the numbers of signal ($\kappa=1$) and background events after the selection cuts. For the four $b$-tagged jets, we also demand $\chi_{HH}<1.6$ which sufficiently suppresses the backgrounds. We end up with 4.8 signal events and 11.7 background events. Based on Eqs.~\ref{eq:dis} and \ref{eq:ex}, we obtain a $5\sigma$ discovery significance and $2\sigma$ exclusion limits on $\kappa$, respectively. It shows that the $5\sigma$ significance requires $\kappa\geq 4.85$ or $\kappa\leq -8.10$, while the $2\sigma$ exclusion bound is $-5.42\leq \kappa\leq 2.16$. The discovery significance of $\lambda_{HH}^{\rm SM}$ is $1.32\sigma$ in the $ZHH$ channel which is comparable to that of the GF channel.

\begin{table}[b]
\caption{The sensitivity to $\lambda_{HHH}=\kappa \lambda_{HHH}^{\rm SM}$ in several production channels of Higgs boson pairs at the HL-LHC. }
\label{tbl:sig}
\begin{tabular}{l|c|c|c}
\hline
 & SM & $5\sigma$ discovery& $2\sigma$  exclusion  \\
 & ($\kappa=1$)  & potential &bound\\
 \hline
 $WHH$ & 1.29$\sigma$  & $\kappa\leq -7.7$, ~$\kappa\geq 4.8$ & $-5.1\leq\kappa\leq 2.2$   \\
 $ZHH$ & $1.32\sigma$   &  $\kappa\leq -8.1$, ~$\kappa\geq 4.8$   & $-5.4\leq\kappa\leq 2.2$  \\
GF($b\bar{b}\gamma\gamma$)~\cite{ATL-PHYS-PUB-2014-019}  & $1.19\sigma$   &  $\kappa\leq -4.5$, ~$\kappa\geq 8.1$  & $-0.2\leq\kappa\leq 4.9$\\
GF($b\bar{b}\gamma\gamma$)~\cite{CMS:2015nat}  & $1.65\sigma$   &  $\kappa\leq -2.6$, ~$\kappa\geq 6.3$  & $~~0.5\leq\kappa\leq 4.1$\\
 VBF~\cite{Dolan:2015zja} & $0.59\sigma$  &  $\kappa\leq -1.7$, ~$\kappa\geq 5.0$  &   $-0.4\leq\kappa\leq 3.5$ \\
 $t\bar{t}HH$~\cite{Englert:2014uqa,Liu:2014rva} & $1.38\sigma$  & $\kappa\leq -11.4, \kappa\geq 6.9$  & $-7.2\leq \kappa\leq 2.5$\\
\hline
\end{tabular}
\end{table}

\section{Discussions and conclusions}

The sensitivity to the trilinear Higgs coupling is summarized in Table~\ref{tbl:sig}. The $\lambda_{HHH}^{\rm SM}$ can be probed at the level of $1.32\sigma$ ($1.29\sigma$) in the $ZHH$ ($WHH$) production at the HL-LHC, respectively. 
The discovery potential of the $VHH$ production is comparable to those of the GF channel given by the ATLAS collaboration ($1.19\sigma$)~\cite{ATL-PHYS-PUB-2014-019} and CMS collaboration ($1.65\sigma$)~\cite{CMS:2015nat} and $t\bar{t}HH$ production ($1.38\sigma$)~\cite{Englert:2014uqa,Liu:2014rva}. The VBF channel can probe $\lambda_{HHH}^{\rm SM}$ only at the level of $0.59\sigma$~\cite{Dolan:2015zja}.
In order to combine all the channels to probe the Higgs trilinear coupling,  we use the following significance formula~\cite{Cowan:2010js},
\beq
\sqrt{-2\log\left(\prod_{i=1}^N\dfrac{L(b_i|s_i+b_i)}{L(s_i+b_i|s_i+b_i)}\right)},
\eeq
where $s_i$ and $b_i$ denotes the number of signal and background events after imposing a set of cuts in the production channel $i$, respectively. The likelihood function is defined as
$L(x_i | n_i)=x_i^{n_i}e^{-x_i}/{n_i!}$. The $\lambda_{HHH}^{\rm SM}$ can be measured at the level of $3.13\sigma$ after combining all the above five channels.

The trilinear Higgs coupling might be modified sizeably in NP models. We vary $\kappa$ to estimate the value of $\lambda_{HHH}$ needed for a $5\sigma$ significance for each $HH$ production channel; see the third column in Table~\ref{tbl:sig}.

\begin{figure}
\includegraphics[scale=0.3]{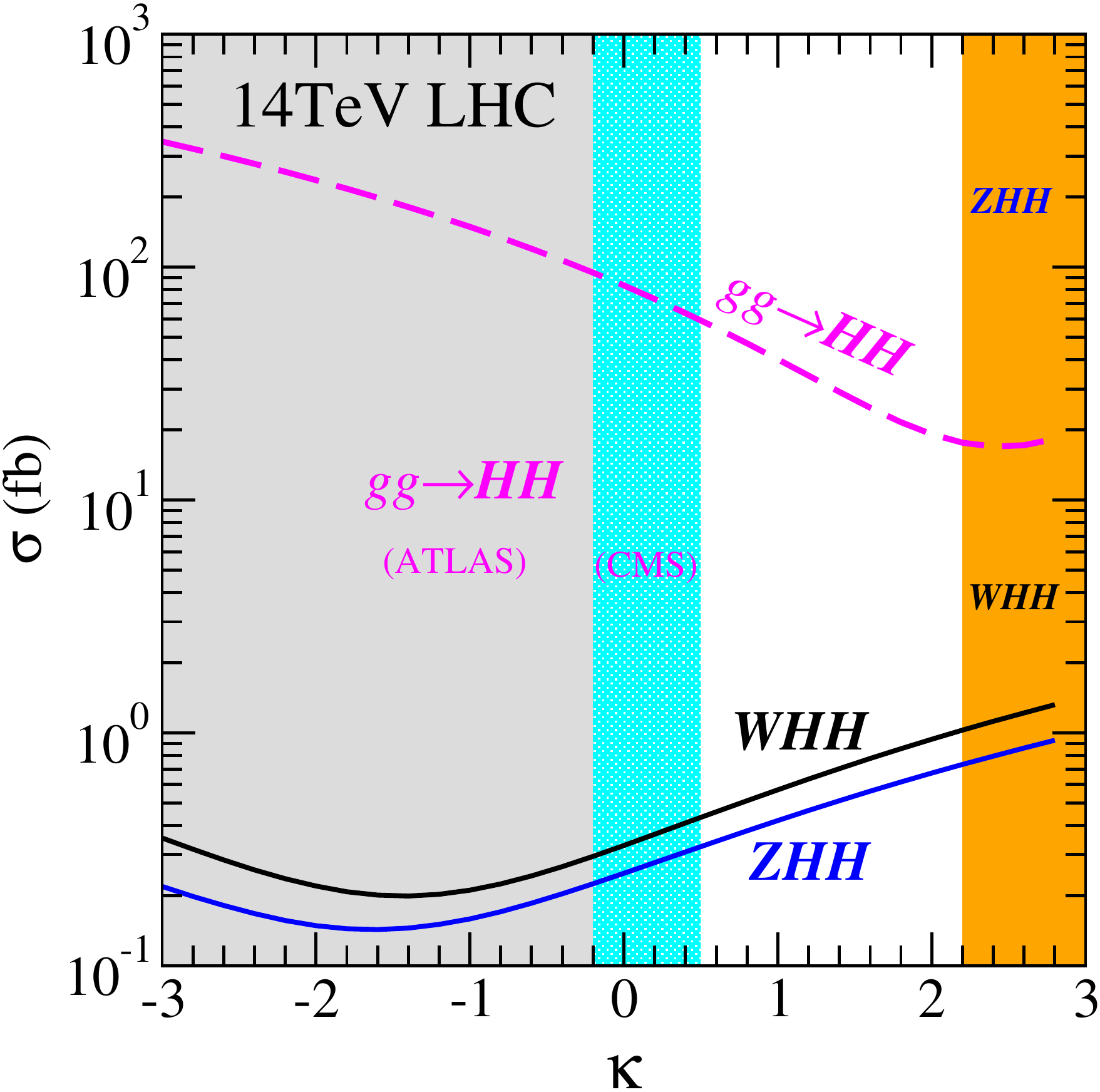}
\caption{The 95\% exclusion bounds on $\lambda_{HHH}=\kappa \lambda_{HHH}^{\rm SM}$ derived from the $VHH$ and GF channels at the HL-LHC.}
\label{fig:sig}
\end{figure}

If no evidence of Higgs pair productions were observed, one could set a $2\sigma$ exclusion limit on $\kappa$ for each $HH$ production channel; see the fourth column in Table~\ref{tbl:sig}.
Figure~\ref{fig:sig} displays the $2\sigma$ exclusion regions on $\kappa$ from the GF channel (gray band from the ATLAS study and cyan band from the CMS result) and from the $WHH$ and $ZHH$ productions (orange band). The most stringent lower limit and upper limit on $\kappa$ arises from the GF channel and the $VHH$ production, respectively, which requires $0.5\leq\kappa\leq 2.2$ at the 95\% confidence level.

\begin{acknowledgements}
The work is supported in part by the National Science Foundation of China under Grand No. 11275009, 11675002 and 11635001.

\end{acknowledgements}

\bibliographystyle{apsrev}
\bibliography{reference}

\end{document}